\documentclass[acmsmall,screen]{acmart}

\usepackage{algorithmic}
\usepackage{graphicx}
\usepackage{textcomp}
\usepackage{xcolor}
\usepackage{xspace}
\usepackage{multirow}
\usepackage{booktabs}
\usepackage{makecell}

\newcommand{\fixme}[1]{\textcolor{orange}{{\it [fixme]}}}

\newcommand{\phead}[1]{\noindent {\bf #1}}

\newcommand{\ihead}[1]{\noindent {\textit {#1}}}

\newcommand{\results}[1]{\noindent {\textbf{\textit{#1}}}}

\usepackage{listings}

\usepackage{xspace}
\newcommand{\tool}{AutoGlue\xspace}

\usepackage{float}

\usepackage{enumitem}

\AtBeginDocument{%
  }

\copyrightyear{2018}
\acmYear{2018}
\acmDOI{XXXXXXX.XXXXXXX}
\acmISBN{978-1-4503-XXXX-X/2018/06}

\begin{document}

\title{Bridging Behavior and Implementation: Automated Java Glue Code Generation for Behavior-Driven Development}


\author{Xinyu Shi}
\affiliation{%
  \institution{University of Alberta}
  \city{Edmonton }
  \country{Canada}}
\email{xshi12@ualberta.ca}

\author{Zhou Yang}
\affiliation{%
  \institution{University of Alberta}
  \city{Edmonton }
  \country{Canada}}
\email{zy25@ualberta.ca}

\author{An Ran Chen}
\affiliation{%
  \institution{University of Alberta}
  \city{Edmonton }
  \country{Canada}}
\email{anran6@ualberta.ca}

\renewcommand{\shortauthors}{Shi et al.}

\begin{abstract}
Behavior-Driven Development (BDD) helps technical and non-technical stakeholders share a common understanding of software requirements through executable natural-language scenarios.
Developers make these scenarios executable by writing glue code, which maps each natural-language step to the corresponding project code. Writing and maintaining glue code requires developers to understand both the intended behavior and the underlying codebase, which makes it one of the most labor-intensive aspects of BDD, especially as requirements evolve.
Although large language models (LLMs) have demonstrated strong code generation capabilities, no prior work has explored automated glue code generation using LLMs. This task requires jointly reasoning over underspecified behavior, BDD artifacts, and large project codebases.
In this paper, we propose \tool, a hierarchical multi-agent framework for automated Java glue code generation.
AutoGlue follows the same behavior-first workflow that developers use in BDD, first understanding the intended behavior, then retrieving relevant project context, and finally generating glue code grounded in both behavior specifications and project code.
\tool separates behavior interpretation, context retrieval, and glue code generation.
In particular, a Behavior Interpreter infers the intent of a step within its scenario context, and a Developer agent coordinates the retrieval of related BDD artifacts and project code before generating glue code.
We evaluate \tool on 1,307 steps from eight open-source Java projects.
\tool improves API F1 by 58.7\% and CodeBLEU by 43.7\% over few-shot prompting.
It also generates directly usable glue code for 46.1\% of the evaluated steps, while most remaining partially correct cases require only minor edits, such as adding missing actions or refining parameters.
Our ablation study further shows that behavior interpretation and project-aware context retrieval each contribute substantially to generation quality.
Overall, our results demonstrate that LLMs can effectively bridge natural-language behavior specifications and project code, opening new opportunities for automating specification-driven software development. We further highlight the limitations of current LLM-based glue code generation and identify promising directions for future research.
\end{abstract}




\maketitle

\section{Introduction}

\begin{figure*}[b]
  \centering
  \includegraphics[width=\textwidth]{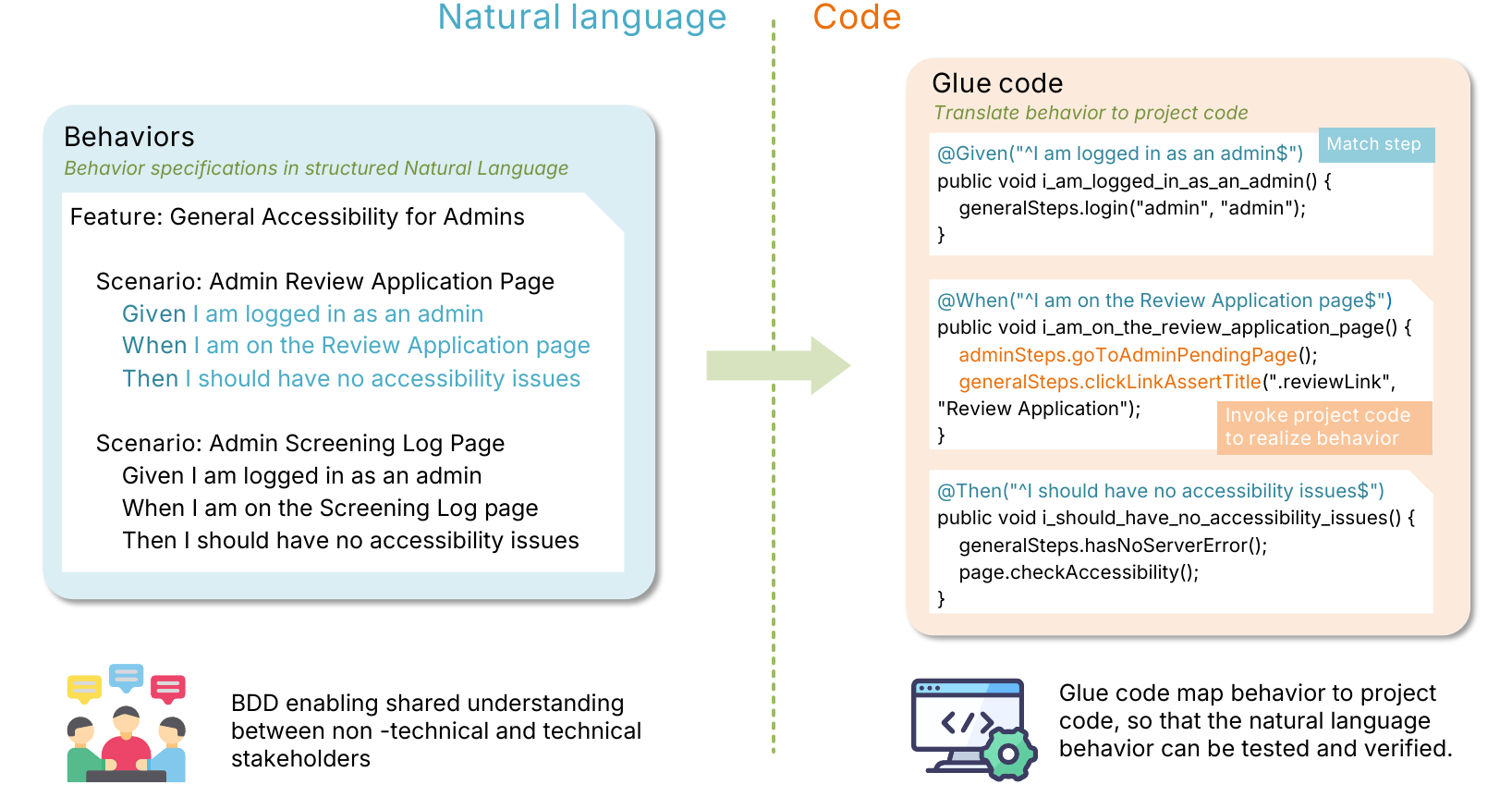}
  \Description{A schematic of a typical BDD workflow, showing feature files and scenarios with steps linked to Java glue code and project code.}
\caption{
BDD describes expected system behaviors as feature files, scenarios, and natural-language steps.
Glue code maps each step to executable logic that invokes the project code, enabling scenarios to run against the system for behavior validation.
}
  \label{fig:overview}
\end{figure*}

Software development requires technical and non-technical stakeholders to maintain a shared understanding of software requirements. In practice, software teams often struggle to stay aligned on what the system is expected to do~\cite{saiedian2000requirements,hofmann2001requirements}. Customers, users, and managers may describe their needs in business terms, while developers require concrete behavioral details for implementation and testing. Prior work in requirements engineering has shown that such communication gaps can lead to misunderstood requirements, weak coordination, quality issues, and wasted effort~\cite{6051639,bjarnason2014challenges}.

Behavior-Driven Development (BDD) addresses this problem by using expected system behavior as a shared artifact between developers and non-technical stakeholders~\cite{smart2023bdd}.
In BDD, the team works together to describe what the system should do before implementation. These specifications are written in natural language, allowing even non-technical stakeholders to read and discuss them, which helps the team develop a shared understanding of the expected behavior.
In practice, BDD specifications are organized into \emph{features} (user goals), \emph{scenarios} (specific cases), and \emph{steps} (context, actions, and expected outcomes introduced by the keywords Given, When, and Then).
For example, the scenario in Figure~\ref{fig:overview} describes an admin reviewing an application, with steps such as ``Given I am logged in as an admin'' and ``Then I should have no accessibility issues''.
These specifications serve not only as documentation but also as executable specifications that verify whether the implementation behaves as expected.
Glue code maps each natural-language step to the corresponding project code, which enables features and scenarios to execute against the system.
By continuously validating that the implementation matches the specified behavior, BDD helps teams maintain a shared understanding of system requirements throughout software development.

However, BDD is costly to adopt in real-world projects~\cite{couto2020understanding, zampetti2020demystifying}, because every step must be wired to the project code by hand through glue code.
As projects evolve, the number of steps can grow to hundreds or even thousands, substantially increasing maintenance effort and technical debt~\cite{zampetti2020demystifying, Shi2026BDDLLM}.
Continual requirement changes further exacerbate this burden, as existing steps and their implementations frequently need to be modified or rewritten.
Moreover, writing glue code requires both an accurate understanding of business logic and deep familiarity with the project codebase, concentrating this work in the hands of a few experienced developers rather than being shared across the team.
Finally, existing techniques~\cite{gao2016generating,kamalakar2013automatically,raharjana2020tool,storer2019behave,zameni2023bdd} offer limited support, as most rely on rule-based mappings or manually maintained artifacts that still require substantial human involvement.

Recent advances in large language models (LLMs) provide an opportunity to automate this process.
Prior work~\cite{hou2024large, zhang2026survey, gao2025current} has shown the potential of LLMs across many software engineering tasks, including code generation~\cite{lin2025robunfr, yang2025empirical}, testing and debugging~\cite{wang2025testeval, shi2025enhancing}, as well as program repair~\cite{bouzenia2024repairagent}.
However, using LLMs to bridge fine-grained steps and project code remains largely unexplored.
In practice, generating glue code poses several challenges.
It requires joint reasoning over multiple artifacts, including individual steps, feature or scenario context, and project code, which makes it more complex than conventional code generation.
In addition, individual steps are often short and underspecified, so their intent can be ambiguous when interpreted in isolation, especially when their behavior depends on other steps in the same scenario or feature.
At the same time, glue code must connect high-level behavior to project code, such as internal functions, domain objects, or project-specific APIs, which is typically scattered across the codebase and requires deep project-specific knowledge.
Moreover, the content of glue code evolves as project behavior changes, while LLMs lack inherent awareness of a specific codebase. Finally, generated glue code must remain consistent with existing implementation styles and conventions, which vary significantly across projects and teams.

To address these limitations, we propose \tool, a hierarchical multi-agent~\cite{he2025llm} framework for automated Java glue code generation from steps.
To the best of our knowledge, \tool is the first approach to automatically generate glue code from steps.
\tool decomposes the process into three sequential stages: behavior interpretation, context retrieval, and glue code generation.
Specifically, a Behavior Interpreter agent first analyzes the target step within its enclosing feature and scenario to derive a description of its intended behavior and its role in the scenario.
A Developer agent then acts as a supervisor~\cite{li2024survey} that coordinates two specialized sub-agents during context retrieval: one retrieves relevant BDD context from similar steps and existing glue code, and the other explores the project codebase to collect project code context.
The multi-agent design is motivated by the complexity of glue code generation, which requires reasoning over both behavior and project code.
\tool separates the reasoning over different types of information into dedicated agents to avoid mixed reasoning, error propagation, and excessive context usage~\cite{guo2024large}.
The hierarchical structure models a behavior-first development process, where behavior is understood before code is written.
The Developer agent uses two specialized sub-agents to retrieve information from different sources and integrate it into glue code, while keeping contexts separate to reduce noise.

To evaluate \tool in real-world practice, we collect 1,307 distinct steps and their corresponding Java glue code from eight open-source GitHub projects.
We choose Java because it is representative of real-world BDD practice.
Java was the first programming ecosystem where BDD was practiced and popularized, supported by early tools such as JBehave~\cite{north2006introducingbdd,jbehave}. Later tools in other languages adopted similar conventions~\cite{zampetti2020demystifying}, including the separation between feature files and project code, as well as pattern-based step binding mechanisms.
Beyond this representativeness, generating Java glue code is harder than for scripting languages.
For instance, Java requires strict annotation binding with regular expressions and correct parameter types, which increases the difficulty of generating valid method signatures~\cite{yang2019predicting}.
In addition, Java projects typically exhibit a strong refactoring culture~\cite{al2015identifying}, clear separation of concerns, and object-oriented designs for glue code.

We conduct a series of experiments to evaluate the effectiveness of our glue code generation framework. Since no established automatic approach exists for glue code generation, we compare our framework with two LLM-based prompting approaches commonly used by developers: few-shot prompting and plain prompting~\cite{marvin2023prompt, jin2024can}.
We first evaluate glue code quality by measuring API overlap, which reflects functional alignment by assessing whether the generated glue code uses the same project APIs as the ground truth, together with similarity metrics~\cite{ren2020codebleu, lin2004rouge, banerjee2005meteor} that capture structural and lexical resemblance to human-written code. Our results show that \tool achieves substantially stronger alignment with ground truth, improving API F1 by over 50\% and CodeBLEU by over 40\% on average compared to prompting-based baselines.
To assess practical usability and to address the limitations of execution-based evaluation caused by complex environments and ambiguous step outcomes in BDD, we employ an LLM-as-a-Judge~\cite{li2024llms, crupi2025effectiveness, wang2025can} to conduct pointwise evaluation against ground-truth glue code.
Our results show that \tool substantially increases the proportion of directly usable glue code, with nearly half of the generated implementations requiring no modification.
Further analysis of partially correct cases shows that these cases typically require only minor changes, most often to add missing actions or to correct imprecise parameters, indicating that the generated glue code is often close to being usable in practice.
We further conduct an ablation study to examine how key design choices affect glue code generation quality.
Our results show that retrieving related BDD artifacts strongly improves similarity to human-written glue code, while access to project code is important for producing correct and directly usable implementations.
Removing behavior interpretation also leads to consistent performance drops across all metrics, indicating that explicit understanding of the step intent provides useful guidance for generation.
In the discussion, we further analyze the efficiency of \tool and discuss its implications, contributions, and remaining open challenges for automated BDD support.
The main contributions of this paper are:

\begin{itemize}[leftmargin=*]
    \item We propose \tool, the first automated approach for generating glue code directly from steps.
    \tool adopts a hierarchical multi-agent design that separates behavior interpretation, context retrieval, and glue code generation.
    
    \item We evaluate \tool on 1,307 step--glue-code pairs collected from eight real-world open-source Java projects.
    \tool consistently outperforms prompting-based baselines in both functional alignment and code similarity.
    
    \item We analyze the usability of generated glue code using an LLM-as-a-Judge.
    Our analysis shows that nearly half of the generated glue code is directly usable, while most partially correct cases require only minor fixes, such as adding missing actions or adjusting parameters.
    
    \item We conduct an ablation study to examine the contribution of key design components.
    We find that grounding LLMs in both behavior specifications and project code substantially improves glue code generation. Future studies are needed to explore how these principles extend to broader specification-to-code automation.
\end{itemize}

\section{Background}

\subsection{Behavior-driven Development}
Translating requirements into working software requires technical and non-technical stakeholders to develop a shared understanding of the expected system behavior. However, stakeholders may interpret natural-language requirements differently because they bring different perspectives, assumptions, and vocabularies. Behavior-driven development (BDD) addresses this problem by using concrete examples written in natural language as a shared basis for discussing expected behavior~\cite{smart2023bdd}. Motivated by test-driven development (TDD), which uses tests to guide software development, BDD uses behavior as a shared reference to guide the development process. It expresses observable system behavior in a form that stakeholders can discuss, refine, and agree upon before connecting it to the implementation. The resulting examples are then formulated as structured behavior specifications and automated as executable checks. In this way, the same examples connect requirements discussion, behavior specification, and implementation verification.

This link is established through an iterative process of Discovery, Formulation, and Automation.
During Discovery, team members discuss concrete examples to clarify expected behavior and identify missing or conflicting interpretations.
During Formulation, the agreed examples are written as structured behavior specifications.
These specifications organize behavior into features, scenarios, and steps.
A feature represents a high-level business capability or user goal.
Each scenario describes one concrete case through a sequence of steps.
\texttt{Given} steps establish the relevant context or preconditions.
\texttt{When} steps describe actions or events.
\texttt{Then} steps state the expected outcomes.
During Automation, glue code connects the steps to project code so that each scenario can be executed.
The three activities form a feedback loop.
If writing or executing an example reveals missing information, the team can return to the discussion and refine its shared understanding.

Figure~\ref{fig:overview} illustrates how a formulated example becomes an executable scenario.
The \texttt{Given} step establishes that the user is logged in as an administrator.
The \texttt{When} step opens the Review Application page, and the \texttt{Then} step checks the page for accessibility issues.
Both scenarios in the figure reuse the login and accessibility steps but open different pages in their \texttt{When} steps.
This reuse allows recurring behavior to be expressed consistently while each scenario retains the context needed to understand the steps.
During execution, glue code connects each natural-language step to a corresponding Java method that calls project code.
The steps run in order to establish the scenario state, perform the action, and check the outcome.
The same scenario can therefore serve as a shared description of expected behavior and as an executable check of the implementation.
Repeated execution also provides feedback when a change no longer satisfies the documented behavior.
These properties help keep requirement discussions, behavior specifications, and project code aligned as the system evolves~\cite{smart2023bdd}.
This benefit depends on maintaining the connection between the steps and project code.
The next subsection explains how glue code establishes and executes this connection.

\subsection{Glue code in BDD}
\label{sec:challenges}

\phead{How glue code works.}
Glue code connects natural-language steps to project code and makes BDD specifications executable.
Before a scenario runs, the patterns in the \texttt{@Given}, \texttt{@When}, and \texttt{@Then} annotations are loaded from the glue code.
Each step is matched against these patterns.
Values captured by a matching pattern are converted to the required types.
They are then passed as parameters to the corresponding Java method.
The Java method is executed and calls the project code needed for the behavior.
For example, in Figure~\ref{fig:overview}, the pattern in the \texttt{@When} annotation matches the step text ``I am on the Review Application page.''
The corresponding Java method calls \texttt{adminSteps.goToAdminPendingPage()} and \texttt{generalSteps.clickLinkAssertTitle(...)}.
These calls open the page and check its title.
Overall, glue code provides an executable connection between behavior specifications and project code.
Developers must maintain this connection as either side evolves.

\begin{figure}
  \centering
  \includegraphics[width=\linewidth]{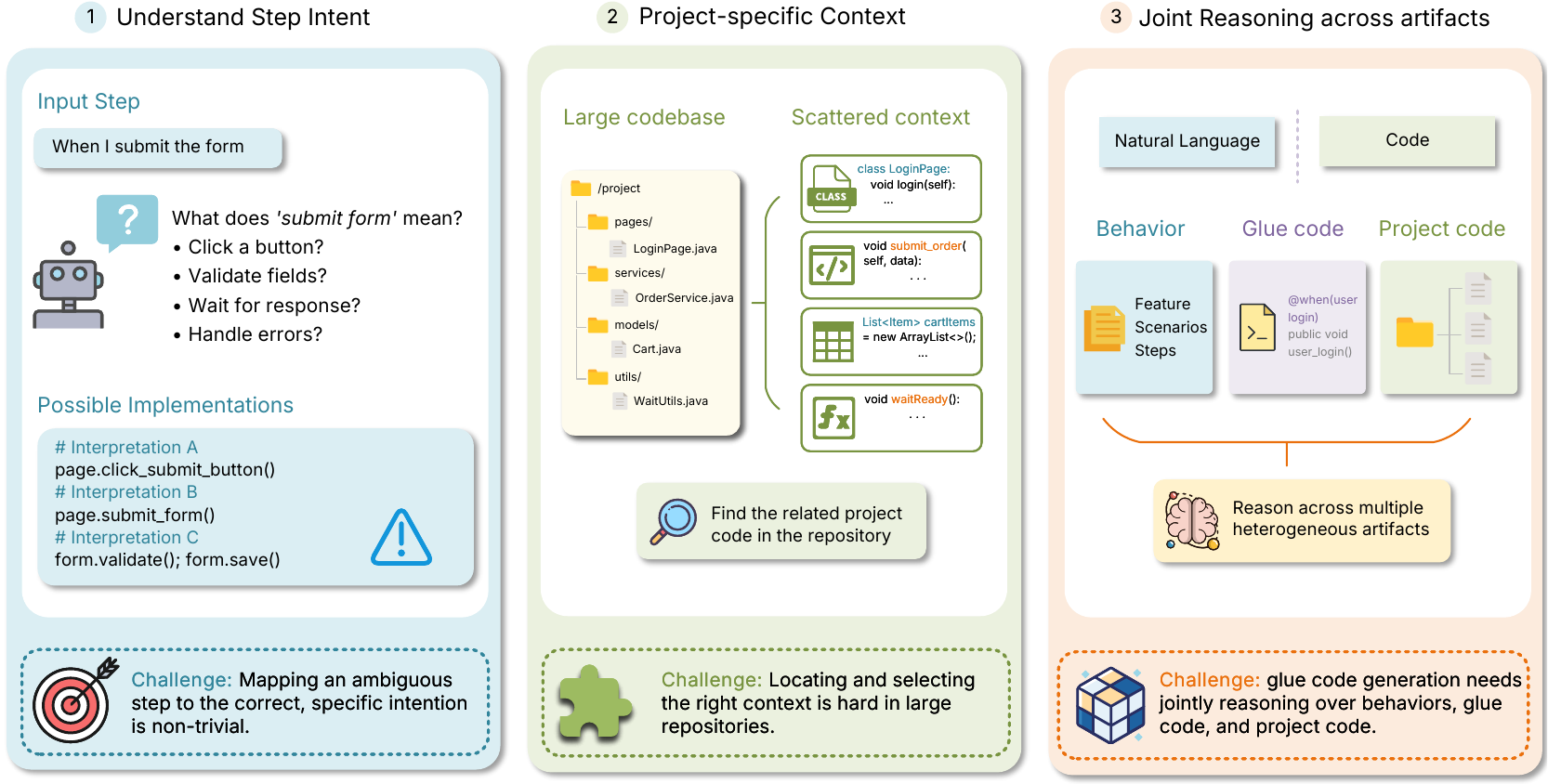}
  \caption{Challenges of glue code generation in BDD.}
  \Description{Overview of the main challenges in BDD glue code generation, including semantic gaps, project-specific context dependence, and joint reasoning over heterogeneous artifacts.}
  \label{fig:glue_code_generation_challenges}
\end{figure}

\phead{Challenges for generation.}
Generating glue code from steps poses several inherent challenges.
(1) Semantic gap between natural-language steps and project code.
Individual steps are usually short and underspecified.
When interpreted in isolation, their intent can be ambiguous, especially when the actual behavior depends on other steps within the same scenario or feature.
For example, a step may simply say that a user submits a form, but the same phrase can have different meanings even within a single project.
The user may refer to different roles or account states, such as a guest, registered customer, administrator, or already authenticated user, and the form may belong to registration, checkout, profile editing, or an administrative workflow.
Each interpretation may require different page objects, validation helpers, domain services, or setup logic.
The glue code must therefore infer which concrete project abstraction should be used from the scenario, feature, and surrounding project code context.
(2) Strong dependence on project- or tool-specific context.
Glue code must connect high-level behavioral descriptions to project code, such as internal functions, domain objects, or project-specific APIs.
This information is often scattered across the codebase and requires detailed project knowledge.
As shown in Figure~\ref{fig:glue_code_generation_challenges}, generation often requires an understanding of the project structure.
The generator must find where the target behavior is implemented in a large codebase.
It may also need to combine objects and methods from different files or functional modules.
(3) Joint reasoning over multiple heterogeneous artifacts.
Correct glue code generation requires reasoning over multiple sources at the same time, including individual steps, scenario or feature context, and project code.
These artifacts use different forms, from natural language to Java code.
The generator must combine them before it can produce glue code that both matches the step and calls the relevant project code.
This makes the task more complex than generating a standalone function and calls for generation strategies designed for BDD.

\phead{Challenges for evaluation.}
Evaluating automatically generated glue code is also non-trivial in the BDD setting.
(1) Lack of execution-based oracles.
Unlike general code generation tasks, execution-based evaluation is difficult to apply to glue code.
There is no clear notion of passing or failing tests that can serve as a reliable oracle, which makes metrics such as pass@k inapplicable.
(2) Absence of step-level outcomes.
Glue code is designed to be executed as part of a complete scenario or feature.
Individual \texttt{Given} or \texttt{When} steps often do not produce observable outcomes or assertions, making it difficult to verify their correctness in isolation.
(3) Complex and implicit environment dependencies.
Glue code is rarely self-contained and its execution depends on project-specific runtime states, such as initialized browsers, connected databases, or preconfigured services.
These dependencies are hard to identify and reconstruct across different open-source projects.

\smallskip
\noindent
\setlength{\fboxsep}{6pt}
\colorbox{gray!10}{%
  \parbox{\dimexpr\linewidth-3\fboxsep\relax}{%
    \textbf{Observation 1:}
    Unlike ordinary project code, glue code is highly context-dependent and requires a deep understanding of the project.
    Glue code generation requires joint reasoning over underspecified natural-language steps, scenario-level behavior, and project code.
  }
}

\section{Related Work}

\subsection{Automated BDD Support }
\label{sec:old_tools}

\phead{BDD adoption and maintenance challenges.}
Prior studies show that BDD is used in practice and can improve communication, requirements understanding, and behavioral validation.
Binamungu et al.~\cite{binamungu2018maintaining} surveyed 75 BDD practitioners from 26 countries and found that BDD is actively used in industry, especially in private organizations.
Practitioners reported that BDD helps teams describe requirements with domain-specific terms, make specifications executable, and help maintenance developers understand the intent of code.
Repository studies also show broad BDD adoption in open-source projects.
Zampetti et al.~\cite{zampetti2020demystifying} studied 50,000 open-source projects and found that 27.25\% use at least one BDD-related framework or tool, with particularly high adoption in Ruby and JavaScript projects.
Similarly, Nascimento et al.~\cite{nascimento2020behavior} interviewed 42 members of agile development teams and found that BDD can improve collaboration and reduce ambiguity in requirements.
Systematic reviews report similar benefits, including customer satisfaction, requirements tracking, artifact reuse, and early validation of expected behavior~\cite{abushama2021effect,farooq2023behavior,arredondo2023benefits}.

However, these benefits come with substantial maintenance costs.
Prior studies report that BDD adoption requires teams to change how they discuss requirements, write scenarios, and maintain executable specifications.
Developers also report a learning curve, difficulty applying BDD correctly, and increased development time and cost~\cite{binamungu2018maintaining,abushama2021effect,arredondo2023benefits}.
As BDD suites grow, specifications become harder to locate, understand, and update, and may contain duplicated scenarios or slow tests~\cite{binamungu2018maintaining}.
Zampetti et al.~\cite{zampetti2020demystifying} also found that developers do not always follow a strict BDD process, with tests often written during or after coding rather than before implementation.
Together, these studies suggest that BDD has clear practical value, but its adoption and long-term use still depend on substantial human effort.
They motivate automated support for reducing BDD maintenance effort, but they mainly study BDD usage, benefits, and challenges rather than generating executable glue code from steps.

\phead{Automated tools for BDD.}
Before the emergence of LLMs, automated support for BDD mainly focused on assisting developers in connecting natural-language scenarios with existing implementations, rather than generating glue code end-to-end.
Most prior tools rely heavily on human involvement, such as manual modeling, predefined rules, or constrained authoring practices.
Several approaches generate BDD artifacts or partial testing code from intermediate representations defined during requirements engineering, requiring developer intervention~\cite{gao2016generating}.
Another line of work maps steps to existing implementation code using probabilistic or rule-based matching techniques.
Such approaches assume that relevant implementation code already exists and therefore do not address glue code generation for new or evolving behaviors~\cite{kamalakar2013automatically,raharjana2020tool,storer2019behave}.
They further rely on manually designed mapping rules, interactive user input, or controlled project settings, which limits their scalability in real-world projects.
More recent tools transform BDD scenarios into formal behavioral models to enable systematic test generation.
While effective for improving test coverage, these approaches rely on explicit formal modeling and do not directly generate executable glue code, leaving the connection between scenarios and project code to developers~\cite{zameni2023bdd}.
Recent LLM-based studies have explored automating parts of the BDD workflow, but their focus remains on behavior discovery rather than glue code generation.
Paduraru et al.~\cite{paduraruagentic} proposed an agent-based approach that assists users in interactively creating and refining BDD scenarios, but only supports reusing existing glue code.
Similarly, Karpurapu et al.~\cite{karpurapu2024comprehensive} evaluated the use of LLMs to generate BDD acceptance tests from requirements with zero- and few-shot prompting.
However, the most labor-intensive component of BDD, the automatic generation of glue code, remains largely unaddressed and continues to rely on manual development.

\subsection{Large Language Models and Agentic Systems}

Large language models (LLMs) are neural language models trained on large-scale text and code corpora to understand and generate natural language, source code, and other structured text~\cite{jin2024llms, vaswani2017attention}.
Given an input prompt, an LLM predicts and generates a response that follows the task description, examples, and contextual information provided in the prompt.
This ability has recently shown strong potential for software engineering tasks.
Existing studies have applied LLMs to a wide range of software engineering tasks, including code generation, test generation, debugging, fault localization, and program repair~\cite{hou2024large,jin2024can,lin2025robunfr,wang2025testeval,shi2025enhancing,bouzenia2024repairagent}.
These tasks often involve understanding what the developer wants to change and producing the corresponding code edits.
This makes LLMs especially relevant to BDD, where expected behavior is explicitly described in natural language and must eventually be connected to executable implementation code.
However, many software engineering tasks cannot be solved reliably from a single isolated prompt.
Generated code often needs to fit the existing codebase, such as its data models, coding style, tests, and framework conventions.
Without sufficient project context, an LLM may produce code that is syntactically plausible but incompatible with the target system.
For this reason, recent LLM-based software engineering approaches increasingly emphasize context-aware generation, retrieval-augmented generation, and repository-level reasoning~\cite{yang2025empirical,shi2025enhancing}.
Instead of generating code only from a short task description, these approaches attempt to ground generation in relevant project artifacts, such as existing source files, tests, documentation, and similar examples.

Agentic systems are systems in which one or more LLM-based agents act toward a goal by following assigned roles, using tools, gathering context, and making intermediate decisions~\cite{jin2024llms, yao2022react}.
They extend LLM-based generation by organizing the model around explicit goals, roles, tools, and decision processes.
Rather than producing one response to one prompt, an agentic system can decompose a task, retrieve information, inspect external artifacts, call tools, and synthesize intermediate results before making a final decision.
Multi-agent systems further divide a complex task among multiple specialized agents, such as planners, retrievers, developers, and reviewers~\cite{li2024survey,guo2024large,he2025llm}. 
Such designs are increasingly explored in software engineering because many development tasks require coordination across heterogeneous artifacts and different forms of reasoning.
At the same time, agentic systems also introduce new challenges, including context noise, error propagation, and coordination overhead.
Therefore, effective agentic designs require clear task decomposition and well-defined responsibilities among agents.
These properties make LLM-based agentic systems a promising fit for BDD glue code generation.
Generating glue code requires understanding the natural-language behavior expressed by a step, interpreting its role within the surrounding scenario, identifying similar existing BDD artifacts, and locating project code that can implement the behavior.
This process naturally involves multiple sources of information and different reasoning needs.
A behavior-oriented component can reduce ambiguity in underspecified steps, while retrieval-oriented components can ground generation in existing BDD conventions and project code.
Thus, an agentic design provides a structured way to bridge natural-language behavior descriptions and project code, which motivates the hierarchical multi-agent framework proposed in this paper.

\section{Methodology}
\label{sec:methodology}



\begin{figure*}[t]
  \centering
  \includegraphics[width=\textwidth]{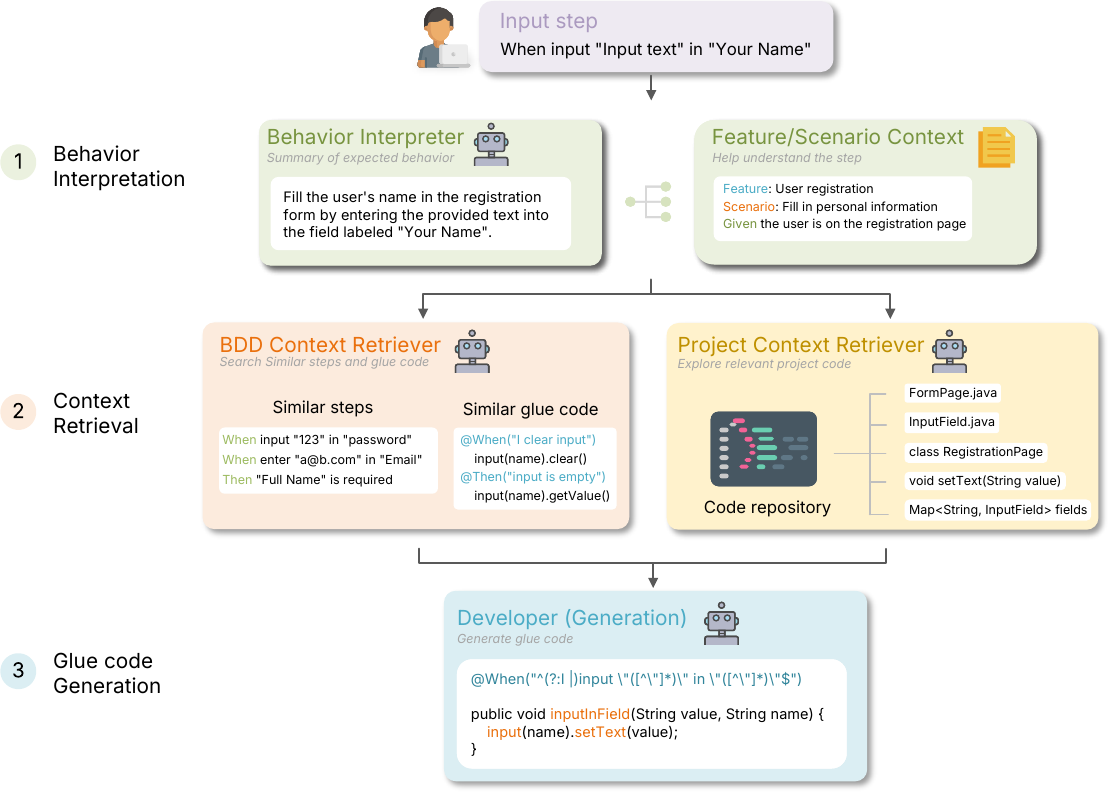}
  \caption{Overview of the \tool framework.}
  \label{fig:methodology}
\end{figure*}

We propose \tool, an LLM-based hierarchical multi-agent framework for generating Java glue code from a target BDD step. As shown in Figure~\ref{fig:methodology}, \tool takes a target BDD step and the project codebase as input, and outputs Java glue code that connects the step to the project code that implements the behavior. The workflow consists of three sequential stages: behavior interpretation, context retrieval, and glue code generation. \tool follows a hierarchical multi-agent design. The Behavior Interpreter agent first clarifies the intended behavior of the target step. The Developer agent then supervises the remaining stages. It delegates context retrieval to two sub-agents, the BDD Context Retriever and the Project Context Retriever, and uses the retrieved context to generate the final glue code. The BDD context helps \tool follow existing glue-code patterns, while the project context helps \tool invoke suitable project code. Using the running example in Figure~\ref{fig:methodology}, given the target step ``When input "Input text" in "Your Name"'', \tool interprets the step, retrieves relevant context, and generates the corresponding Java glue code.

\subsection{Behavior Interpretation}

Individual BDD steps are often underspecified when interpreted in isolation. The same step may correspond to different project behaviors depending on the surrounding scenario and feature. \tool therefore begins by interpreting the intended behavior before retrieving project context.
The Behavior Interpreter agent performs this stage by starting from the target step and using its feature and scenario context to clarify information that may be implicit in the step text. For example, in Figure~\ref{fig:methodology}, the target step already indicates an input action, but the surrounding context shows that it occurs on a user registration page and that the field ``Your Name'' refers to the user's name in the registration form. Based on this context, the Behavior Interpreter agent produces a behavior summary that states what the glue code should accomplish, without specifying how the code should be implemented. By producing this summary before retrieval and generation, \tool reduces ambiguity in the target step and provides clearer guidance for the later stages.

\subsection{Context Retrieval}

Once \tool identifies the intended behavior, it gathers the contextual information needed to implement it. Since glue code must align with both existing BDD artifacts and the project codebase, the Developer agent delegates this retrieval step to two specialized sub-agents.
Before writing the final code, the Developer agent invokes two retrieval sub-agents. The BDD Context Retriever collects context from existing BDD artifacts, while the Project Context Retriever collects context from the project codebase. These two sub-agents gather and organize useful information for the Developer agent, but do not generate the final glue code.

\ihead{The BDD Context Retriever} sub-agent searches for steps that are semantically similar to the target step and examines their corresponding glue code using embedding-based retrieval with cosine similarity. It can also directly search within existing glue code to identify potentially relevant implementations. These retrieved steps often involve similar domain objects or operations, and allow the framework to identify established behavioral patterns, naming conventions, and implementation styles already present in the project.
The retrieved results are filtered and consolidated before being returned to the Developer agent, providing references that help ground subsequent code generation in existing BDD practices within the project. The retriever selectively aggregates the most relevant steps and code snippets as useful references.

\ihead{The Project Context Retriever} sub-agent is responsible for collecting project context required for implementing the glue code. It explores the project codebase by examining the project file structure, identifying potentially relevant source files, inspecting symbol trees such as classes and methods, and selectively checking code snippets that may be involved in implementing the target step. These may include domain objects, helper utilities, or existing test infrastructure that the glue code is expected to interact with. This design is particularly useful when no closely related existing glue code is available as a direct reference, as the implementation must rely more heavily on project code rather than prior examples.
The retrieved information is consolidated and returned to the Developer agent, enabling context-aware glue code generation.

\subsection{Glue Code Generation}

With the relevant context from both BDD artifacts and the project codebase collected, \tool proceeds to the final stage of glue code generation. In this stage, the Developer agent jointly reasons over the target step, the behavior summary, and the retrieved context to generate the final Java glue code.
The output includes the step annotation, method signature, parameters, and method body. The generated code should match the natural-language target step and invoke suitable project code. In this way, \tool separates behavior understanding, context retrieval, and code generation, while allowing the Developer agent to make the final generation decision based on the collected evidence.

\section{Data Collection}

\begin{table*}[tb]
\centering
\caption{Collected open-source Java BDD repositories and statistics of BDD artifacts}
\label{tab:bdd_repos}
\small
\begin{tabular}{lr|rrr|r}
\toprule
\textbf{Repository} & \textbf{LOC} & \textbf{Features} & \textbf{Scenarios} & \textbf{Steps} & \textbf{Step--glue-code pairs} \\
\midrule
metasfresh                 & 3,408,473 & 191 & 758 & 12,039 & 615 \\
CxFlow                  & 89,636   & 45  & 159 & 708   & 309 \\
PSM                     & 48,173   & 27  & 154 & 706   & 140 \\
SYMON                   & 28,789   & 19  & 55  & 4,711  & 104 \\
JDI Light                 & 139,214  & 36  & 264 & 696   & 89  \\
SpringMVC Router              & 2,125    & 5   & 44  & 136   & 25  \\
Worblehat                   & 2,595    & 7   & 12  & 51    & 20  \\
Datadog Java APM                & 375,540  & 35  & 43  & 91    & 5   \\
\bottomrule
\end{tabular}
\end{table*}

\phead{Studied systems.}
To study BDD glue code generation in the wild, we construct an evaluation dataset from open-source Java projects that adopt BDD. Our goal is to collect realistic pairs of steps and their corresponding glue code, which together form the basis for automated glue code generation.
We first identify candidate Java repositories on GitHub that adopt BDD practices. Following prior study~\cite{chandorkar2022exploratory}, we treat the presence of BDD feature files (\texttt{.feature}) as a necessary indicator of BDD adoption, as these files contain executable behavioral specifications.
For each project, we further verify the existence of Java glue code using BDD annotations (\texttt{@Given}, \texttt{@When}, \texttt{@Then}), ensuring that the feature-file specifications are actively bound to executable test code.
To ensure that the collected data reflects realistic BDD usage, we restrict our dataset to projects that are non-trivial in scale, actively maintained and well-documented. Based on these criteria, we identify eight open-source Java projects spanning diverse application domains.
We align steps with their corresponding glue code using pattern matching to ensure that each aligned pair reflects an executable relationship in the original project.
To ensure a clean and non-redundant evaluation dataset, we apply the following filtering rules.
(1) First, we remove steps that have no corresponding glue code.
This is common in real-world projects that are still under development, where scenarios are defined but the business logic has not yet been implemented.
(2) Second, multiple steps may map to the same glue code method, differing only in parameters.
To avoid redundancy and prevent biased generation instances, we randomly select one pair for each such group.
As shown in Table~\ref{tab:bdd_repos}, after filtering and alignment, the final dataset contains 1,307 step--glue-code pairs across the eight open-source projects. Table~\ref{tab:bdd_repos} summarizes the projects included in our study and reports key statistics, such as the number of feature files, steps, and the filtered step--glue-code pairs per project.

\phead{Empirical observations.}
We further analyze the collected data to understand how BDD artifacts are used in practice.
On average, each scenario contains three \texttt{Given} steps, two \texttt{When} steps, and three \texttt{Then} steps.
We examine the evolution of system behavior.
Among all scenarios, 49.2\% of the steps are created at the same time as their corresponding feature files, while the remaining 50.8\% are added at a later stage.
In addition, 65.1\% of scenarios are modified after their initial creation, suggesting that behavioral specifications often evolve over time.
We also examine the complexity and evolution of the glue code.
Compared to core business logic, glue code methods are structurally lightweight, with a mean cyclomatic complexity of 1.5, an average of seven lines of code, and fewer than one parameter (0.92) per method on average.
Our evolution study shows that glue code evolves over time.
Each glue code method is modified four times on average during project evolution, reflecting ongoing changes in both behavioral specifications and implementation details.

\smallskip
\noindent
\setlength{\fboxsep}{6pt}
\colorbox{gray!10}{%
  \parbox{\dimexpr\linewidth-3\fboxsep\relax}{%
    \textbf{Observation 2:}
    Both steps and their implementations evolve frequently, necessitating a more adaptive and automation-friendly approach to glue code generation and maintenance.
  }
}

\section{Experimental Settings}


\subsection{Baseline Selection}
Automated generation of Java glue code from natural-language steps remains an underexplored problem.
As discussed in Section~\ref{sec:old_tools}, prior solutions rely on human-defined mappings between steps and project APIs and do not support direct code generation, which makes them unsuitable as baselines for comparison.
Meanwhile, LLMs are increasingly adopted in practice as general-purpose assistants for code generation in software development workflows~\cite{jin2024can,lin2024soen}.
Under such circumstances, new baselines are needed to reflect real-world LLM usage.
To approximate real-world settings, two prompt-based LLM baselines are constructed to represent common ways developers interact with LLMs during BDD development.
Both baselines generate glue code directly from the input step.
The two designed baselines are as follows:

\begin{itemize}[leftmargin=*]
  \item \textbf{Few-shot prompting} provides the LLM with three example pairs of steps and their corresponding glue code implementations. Examples are randomly sampled from the same project as the target step, ensuring that they are drawn from the same codebase without manual selection. Such in-context examples allow the model to observe concrete mappings between step descriptions and glue code implementations, which is a commonly adopted strategy to improve code generation performance~\cite{marvin2023prompt}. The target step is then appended to the prompt, and the model is instructed to generate the corresponding glue code following the given examples.

  \item \textbf{Plain prompting} represents a minimal interaction setting that is also commonly adopted by developers. The model receives only a natural-language instruction describing the task, together with the target step. No in-context examples or project-specific information are provided. This baseline captures zero-shot generation behavior and serves as a lower-bound reference.
\end{itemize}

\subsection{Evaluation Metrics}
To comprehensively assess the quality of generated glue code, we design a set of evaluation metrics that measure functional alignment, code similarity, and practical usability.

\phead{Functional alignment.}
To measure the functional alignment between generated glue code and the intended behavior, we design a set of API overlap metrics that compare the usage of APIs in generated glue code against those in corresponding human-written glue code.
In this work, we define the APIs as the program elements in glue code, including method calls and object references, which represent the invocation of corresponding functional modules.
We identify the program elements through static analysis based on AST parsing. Specifically, we parse both generated and ground truth Java glue code into ASTs and extract method invocations and object references. API overlap is then computed by exact symbol matching between generated and reference glue code, which precisely reflects whether the glue code links to the required functional elements.

\begin{itemize}[leftmargin=*]
  \item \textbf{API Precision} measures the proportion of program elements in the generated glue code that also appear in the corresponding human-written implementation, computed as $\frac{|A_{\text{gen}} \cap A_{\text{ref}}|}{|A_{\text{gen}}|}$, where $A_{\text{gen}}$ and $A_{\text{ref}}$ denote the sets of APIs extracted from the generated and reference glue code, respectively. This metric characterizes how accurately the model selects relevant project-level APIs without introducing unnecessary or spurious calls.

  \item \textbf{API Recall} captures the extent to which the generated glue code recovers the program elements present in the reference implementation, defined as $\frac{|A_{\text{gen}} \cap A_{\text{ref}}|}{|A_{\text{ref}}|}$. It reflects the model's ability to cover the essential elements required to realize the intended step behavior.

  \item \textbf{API F1} balances correctness and coverage by computing the harmonic mean of API Precision and API Recall: $\frac{2 \cdot \mathrm{Precision} \cdot \mathrm{Recall}}{\mathrm{Precision} + \mathrm{Recall}}$.
\end{itemize}

\phead{Code similarity.}
To measure lexical and structural similarity between generated and human-written glue code, we further adopt widely used code similarity metrics.
These metrics provide a complementary perspective by characterizing similarities in code representation, organization, and token-level patterns, independent of functional correctness.
Specifically, we consider the following widely used automatic metrics~\cite{ouedraogo2025beyond, paul2024benchmarks} to assess lexical and structural similarity.

\begin{itemize}[leftmargin=*]
  \item \textbf{CodeBLEU}~\cite{ren2020codebleu} is a composite code similarity metric that integrates n-gram overlap, syntax matching based on abstract syntax trees, and data-flow consistency, capturing both surface-level token similarity and higher-level correspondence in generated code.

  \item \textbf{METEOR}~\cite{banerjee2005meteor} measures lexical similarity through exact and partial token matches, making it more tolerant to minor variations in naming and formatting and suitable for comparing semantically similar but syntactically different implementations.

  \item \textbf{ROUGE-L}~\cite{lin2004rouge} evaluates sequence-level similarity by computing the longest common subsequence between generated and reference implementations, thereby reflecting alignment in code ordering and sequential structure.
\end{itemize}

Taken together, this metric suite offers a coarse-grained but multi-faceted assessment of glue code quality, covering both project-level API grounding and form-level implementation similarity.

\phead{Usability.}
To address evaluation challenges, we evaluate the usability of generated glue code using an output-based, reference-aware LLM-as-a-Judge framework.
Recent studies have shown that output-based LLM judges, which directly reason over the generated artifact and emit an explicit judgment, achieve higher alignment with human developers in software engineering tasks~\cite{wang2025can,crupi2025effectiveness}.
These approaches are particularly suitable when execution-based validation is infeasible and when correctness, rather than surface similarity, is the primary concern.
In our setting, the availability of human-written ground-truth glue code enables direct comparison between generated and reference implementations, which makes the reference-aware LLM-as-a-Judge evaluation particularly suitable.
Following the formal characterization of LLM-as-a-Judge in prior work~\cite{he2025code}, we adopt a \emph{pointwise} evaluation setting.
Each generated glue code is assessed independently, without comparison to alternative candidates.
For each instance, the judge is provided with three inputs: the step input, the generated glue code, and the corresponding ground truth human-written glue code as a reference.
While reference-free judging has been explored in prior work, multiple studies indicate that reference-aware evaluation can better anchor semantic reasoning and reduce false positives caused by superficially plausible but incorrect code~\cite{crupi2025effectiveness,fandina2025automated}.

The LLM judge is instructed to produce a single holistic assessment reflecting whether the generated glue code constitutes an acceptable implementation of the step within the given project context.
Rather than decomposing the evaluation into multiple independent sub-scores, we follow evidence that holistic judgments more closely resemble human evaluation behavior and yield more stable score distributions~\cite{wang2025can}.
To guide the model's reasoning and mitigate overestimation, we explicitly require the judge to consider multiple criteria during its internal deliberation, including semantic equivalence to the ground-truth behavior, correctness of step-to-code binding, consistency with project-specific API usage, and the presence of missing or incorrect logic.

These criteria guide how the LLM reasons about the glue code and help achieve more stable and accurate judgments.
The final judgment is expressed in two complementary forms.
First, the judge assigns a continuous semantic correctness score in the range $[1.0, 5.0]$, allowing fine-grained distinctions among imperfect implementations, as advocated by prior empirical studies on LLM-based evaluators ~\cite{wang2025can}.
Second, the score is mapped to a categorical label using a predefined three-level scheme.
This hybrid design combines the sensitivity of continuous scoring with the interpretability and reporting convenience of discrete labels, a practice commonly adopted in recent LLM-as-a-Judge frameworks~\cite{crupi2025effectiveness,he2025code}.
We define the categories in the prompt as follows:
\begin{itemize}[leftmargin=*]
    \item \textbf{Exact Match} corresponds to implementations that can be \textbf{directly accepted} without modification, as they correctly realize the semantics of the step, preserve the intended behavior of the reference implementation, and use appropriate project APIs.
    \item \textbf{Partial Match} describes implementations that require \textbf{minor fixes} to become acceptable. These implementations capture the main intent of the step but exhibit secondary deviations, such as incomplete condition handling, minor behavioral inconsistencies, or limited API misuse.
    \item \textbf{Mismatch} refers to implementations that are \textbf{not usable} in practice, as they fail to satisfy the intended behavior, contain substantial semantic errors or irrelevant logic, or omit core functionality.
\end{itemize}

Importantly, this categorization is grounded in practical usability rather than strict textual or structural similarity.
As emphasized in prior work, the ultimate goal of LLM-based evaluation in software engineering is to approximate how a competent developer would judge the usefulness and correctness of a generated artifact in context~\cite{he2025code,wang2025can}.
The prompt for the LLM judge is shown below:

\begin{center}
  \includegraphics[width=0.85\linewidth]{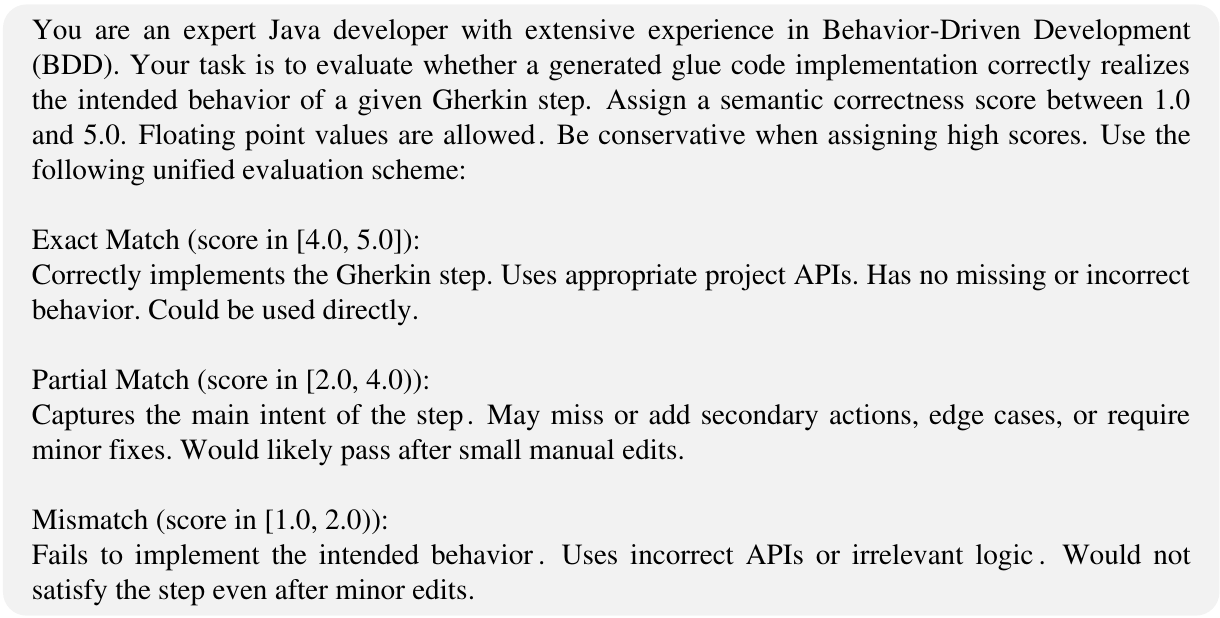}
\end{center}

\subsection{Implementation and Environment}
All experiments are conducted using the same LLM across all approaches.
We employ \texttt{GPT-5 mini}~\cite{openaiGPT5Mini} for all LLM components, which is a recently released model, balancing capability and computational cost.
For embedding-based retrieval, we employ the \texttt{text-embedding-3-small} model provided by OpenAI.
When conducting experiments with \tool, we use the provider's default temperature setting, while setting the temperature to zero for the LLM-as-a-Judge evaluation to ensure stability and consistency.
\tool is organized into a hierarchical multi-agent framework implemented using \emph{LangGraph}, consisting of a Behavior Interpreter, a Developer agent, and two retrieval sub-agents, as described in Section~\ref{sec:methodology}.
Prompt designs are kept concise to reduce unnecessary token consumption and to reserve sufficient context.
To ensure fair comparison, all methods are evaluated under the same model configuration and experimental settings.
To prevent data leakage, ground-truth glue code is strictly excluded during generation.
Since multiple steps can map to the same glue code, we apply different controls at different stages.
During retrieval, we exclude all steps aligned with the target glue code.
During behavior interpretation, we retain the target step but remove access to its glue code, preventing information leakage while allowing behavior analysis.
Similarly, during project context retrieval, the target glue code is removed from the project code.
As a result, the framework has no access to ground truth during the generation task.

\section{Results}
\noindent This section presents the results of our experiments by proposing and answering the following research questions (RQs). 

\subsection{RQ1: How effective is \tool at generating glue code in the wild?}
\label{sec:RQ1}

\phead{Motivation.}
In BDD, glue code connects steps to project code.
We evaluate whether \tool can generate glue code that is functionally aligned and consistent with human-written implementations under real-world settings.

\phead{Approach.}
To assess the effectiveness of \tool in realistic, in-the-wild settings, we evaluate the generated glue code by directly comparing it against human-written implementations (ground truth) collected from open-source Java BDD projects.
For each step in our dataset, \tool generates a corresponding glue code implementation under the default and fixed configurations, without project-specific tuning or manual post-processing. This reflects a practical usage scenario in which developers apply the tool out of the box to evolving real-world codebases.
To evaluate different aspects of glue code quality, we employ a combination of complementary automatic metrics.
In this section, we use functional alignment (API overlap) and code similarity metrics to evaluate the average performance of the generated glue code.
These automatic metrics enable scalable evaluation across a diverse set of real-world projects and establish a quantitative baseline for the more fine-grained semantic analysis conducted in RQ2.
For the API overlap evaluation, we report precision, recall, and F1 score using macro averaging, where these metrics are first computed independently for each step and then averaged across all steps to obtain per-project and overall scores.
In addition, we validate step-to-glue-code binding by testing whether the annotation patterns in the generated glue code successfully match the target steps.

\begin{table}[tb]
\small
\caption{Per-project and overall functional alignment (API overlap) comparison}
\begin{center}
\resizebox{\linewidth}{!}{%
\begin{tabular}{l|ccc|ccc|ccc}
\toprule
\multirow{2}{*}{\textbf{Project}}
& \multicolumn{3}{c|}{\textbf{\tool}}
& \multicolumn{3}{c|}{\textbf{Few-shot}}
& \multicolumn{3}{c}{\textbf{Plain Prompt}} \\

& Precision & Recall & F1
& Precision & Recall & F1
& Precision & Recall & F1 \\
\midrule
metasfresh        & 0.629 & 0.694 & 0.626 & 0.469 & 0.434 & 0.396 & 0.291 & 0.359 & 0.287 \\
CxFlow            & 0.517 & 0.574 & 0.514 & 0.395 & 0.330 & 0.319 & 0.365 & 0.316 & 0.281 \\
PSM               & 0.891 & 0.891 & 0.887 & 0.628 & 0.573 & 0.590 & 0.251 & 0.389 & 0.258 \\
SYMON             & 0.720 & 0.827 & 0.748 & 0.488 & 0.450 & 0.437 & 0.401 & 0.334 & 0.320 \\
JDI Light         & 0.873 & 0.881 & 0.872 & 0.523 & 0.583 & 0.538 & 0.187 & 0.421 & 0.238 \\
SpringMVC Router  & 0.858 & 0.864 & 0.842 & 0.615 & 0.568 & 0.541 & 0.415 & 0.344 & 0.335 \\
Worblehat         & 0.660 & 0.689 & 0.649 & 0.392 & 0.457 & 0.405 & 0.368 & 0.309 & 0.320 \\
Datadog Java APM  & 0.857 & 0.897 & 0.874 & 0.629 & 0.731 & 0.662 & 0.407 & 0.452 & 0.415 \\
\midrule
\textbf{Overall}  & \textbf{0.660} & \textbf{0.714} & \textbf{0.660}
                  & 0.476 & 0.440 & 0.416
                  & 0.310 & 0.353 & 0.284 \\
\bottomrule
\end{tabular}
}
\label{tab:api-metrics}
\end{center}
\end{table}

\begin{table}[tb]
\small
\centering
\caption{Per-project and overall glue code similarity comparison}
\begin{center}
\resizebox{\linewidth}{!}{%
\begin{tabular}{l|ccc|ccc|ccc}
\toprule
\multirow{2}{*}{\textbf{Project}}
& \multicolumn{3}{c|}{\textbf{\tool}}
& \multicolumn{3}{c|}{\textbf{Few-shot}}
& \multicolumn{3}{c}{\textbf{Plain Prompt}} \\

& CBLEU & METEOR & RL
& CBLEU & METEOR & RL
& CBLEU & METEOR & RL \\
\midrule
metasfresh        & 0.585 & 0.604 & 0.573 & 0.372 & 0.365 & 0.349 & 0.348 & 0.321 & 0.236 \\
CxFlow            & 0.530 & 0.541 & 0.461 & 0.397 & 0.357 & 0.310 & 0.427 & 0.318 & 0.241 \\
PSM               & 0.900 & 0.910 & 0.897 & 0.745 & 0.724 & 0.653 & 0.581 & 0.488 & 0.260 \\
SYMON             & 0.737 & 0.735 & 0.691 & 0.472 & 0.471 & 0.442 & 0.439 & 0.393 & 0.281 \\
JDI Light         & 0.851 & 0.870 & 0.847 & 0.599 & 0.634 & 0.539 & 0.518 & 0.343 & 0.193 \\
SpringMVC Router  & 0.751 & 0.800 & 0.766 & 0.459 & 0.488 & 0.471 & 0.401 & 0.371 & 0.279 \\
Worblehat         & 0.600 & 0.612 & 0.568 & 0.498 & 0.463 & 0.376 & 0.417 & 0.354 & 0.275 \\
Datadog Java APM  & 0.718 & 0.902 & 0.852 & 0.481 & 0.669 & 0.621 & 0.362 & 0.486 & 0.463 \\
\midrule
\textbf{Overall}  & \textbf{0.641} & \textbf{0.655} & \textbf{0.614}
                  & 0.446 & 0.433 & 0.396
                  & 0.412 & 0.348 & 0.243 \\
\bottomrule
\end{tabular}
}
\label{tab:codegen-metrics}
\end{center}
\end{table}

\phead{Results.}
\results{\tool improves average API F1 by 58.7\% over few-shot prompting baseline and by 132.4\% over plain prompting baseline, showing markedly stronger functional alignment.}
Table~\ref{tab:api-metrics} shows that \tool reaches an overall API F1 score of 0.660, compared to 0.416 for few-shot prompting and 0.284 for plain prompting.
Both API Precision and API Recall contribute to this gain across all evaluated projects.
In particular, the overall API Recall of \tool reaches 0.714, which is 27.4 percentage points higher than few-shot prompting and more than twice that of plain prompting, while its API Precision reaches 0.660, exceeding few-shot prompting by 38.7\% and plain prompting by 112.9\%.
This pattern suggests that \tool recovers a larger fraction of essential project-level API interactions needed to realize the intended behavior of steps, instead of relying on generic or weakly related calls.

\results{\tool improves average CodeBLEU, METEOR, and ROUGE-L by 43.7\%, 51.3\%, and 55.1\% over few-shot prompting baseline, indicating closer alignment with human-written glue code.}
As shown in Table~\ref{tab:codegen-metrics}, \tool achieves overall CodeBLEU, METEOR, and ROUGE-L scores of 0.641, 0.655, and 0.614.
The corresponding scores for few-shot prompting are 0.446, 0.433, and 0.396, while plain prompting yields consistently lower values of 0.412, 0.348, and 0.243.
The largest relative improvement appears in ROUGE-L, reflecting stronger sequence-level alignment between generated and reference implementations.
At the same time, gains in CodeBLEU and METEOR point to improved alignment in syntactic patterns and token-level realization.

\noindent We also check the correctness of annotation pattern matching for all generated glue code. No pattern-matching errors are observed for \tool or for either prompt-based baseline, indicating that LLMs are capable of correctly handling step-to-glue-code binding in practice.

\smallskip
\noindent
\setlength{\fboxsep}{6pt}
\colorbox{gray!10}{%
  \parbox{\dimexpr\linewidth-3\fboxsep\relax}{%
    \textbf{RQ1 Takeaway:}
    \tool can generate high-quality Java glue code in real-world projects, achieving significantly better functional alignment and code similarity than prompt-based baselines, while maintaining correct step-to-glue-code binding.
  }
}

\subsection{RQ2: How usable is the generated glue code?}
\label{sec:RQ2}



\phead{Motivation.}
While automatic metrics such as API overlap and code-similarity measures provide useful signals about the average quality of generated glue code, they do not directly reflect whether the generated code can be practically adopted.
In real-world BDD projects, the usability of glue code is primarily determined by whether it correctly implements the intended step behavior and can be integrated into existing test suites with little or no modification.

\phead{Approach.}
Evaluating the usability of generated glue code is challenging in BDD, as execution-based oracles are unavailable, individual steps rarely have observable outcomes, and glue code is not self-contained.
To address this, we adopt a pointwise, reference-aware LLM-as-a-Judge framework that directly compares generated glue code against the ground truth and produces usability judgments.
As discussed in Section~\ref{sec:challenges}, we prompt the LLM judge to assign a usability score in the range $[1.0, 5.0]$, where each score level is explicitly defined with a corresponding category.
The generated glue code is categorized into \emph{Exact Match}, \emph{Partial Match}, and \emph{Mismatch}, corresponding to cases that require no modification, minor modification, or are unusable, respectively.
To validate the accuracy of the LLM judge, we randomly sample 120 cases and conduct consistency and correlation analyses against human evaluation using Spearman’s rank correlation coefficient and Cohen's $\kappa$.

To better reflect practical usability, we further analyze the \textit{Partial Match} cases.
These cases require manual modification but are not completely unusable, and therefore represent the main source of post-editing effort in practice.
As a result, \textit{Partial Match} serves as the key boundary for estimating the manual effort needed to make the generated glue code usable.
We analyze the dominant issue in Partial Match cases that prevents the generated glue code from being directly acceptable.
We adopt a single-label strategy that mirrors how developers typically reason about fixes by identifying the most dominant issue to be addressed first, which is consistent with prior findings that LLM-based evaluators are more stable and interpretable when providing a single primary judgment~\cite{fandina2025automated}.
To derive these subcategories, we manually inspect a random sample of 100 Partial Match instances drawn across all evaluated projects and identify a set of recurring correction patterns.
Based on this observation, we define the following categories and instruct the LLM judge to output the corresponding classification in the prompt:

\begin{itemize}[leftmargin=*]
  \item \textbf{Missing Action} describes implementations that capture the overall intent of the step but omit one or more required actions necessary to fully realize the intended behavior.

  \item \textbf{Parameter Mismatch} characterizes implementations that use appropriate project APIs but supply incorrect, incomplete, or imprecise parameters, resulting in behavior that deviates from the intended semantics.

  \item \textbf{Condition Mismatch} applies to cases where the main action is present, but the conditional checks or branching logic do not correctly reflect the execution conditions.

  \item \textbf{Over Implementation} denotes implementations that introduce additional behaviors, API calls, or control logic that are not required by the step or the reference implementation.

  \item \textbf{Invocation Order Mismatch} refers to implementations that invoke relevant operations in an incorrect or suboptimal order, leading to partially correct or unintended behavior.
\end{itemize}

These subcategories are defined in terms of observable behavioral deviations rather than syntactic properties.
They are intended to capture developer-relevant correction patterns in partially correct glue code, providing an interpretable characterization of why an implementation is not directly acceptable.
In addition, this categorization offers a structured basis for analyzing and comparing Partial Match cases in future studies on BDD glue code generation.

\begin{figure*}[t]
    \centering
    

\captionof{table}{Results of LLM-as-a-Judge evaluation of \tool and baseline methods on 1,307 glue code instances}
\label{tab:rq2_usability}

\begin{tabular}{lccc}
\toprule
\textbf{Method} & \textbf{Exact Match} & \textbf{Partial Match} & \textbf{Mismatch} \\
\midrule
\tool                   & 603  & 457  & 247  \\
Few-shot Prompting      & 289  & 517  & 501  \\
Plain Prompting         & 175  & 437  & 695  \\
\bottomrule
\end{tabular}

    \bigskip
    
    \Description{Bar chart comparing Exact Match, Partial Match, and Mismatch outcomes across \tool, few-shot, and plain-prompting baselines.}
    \includegraphics[width=0.85\textwidth]{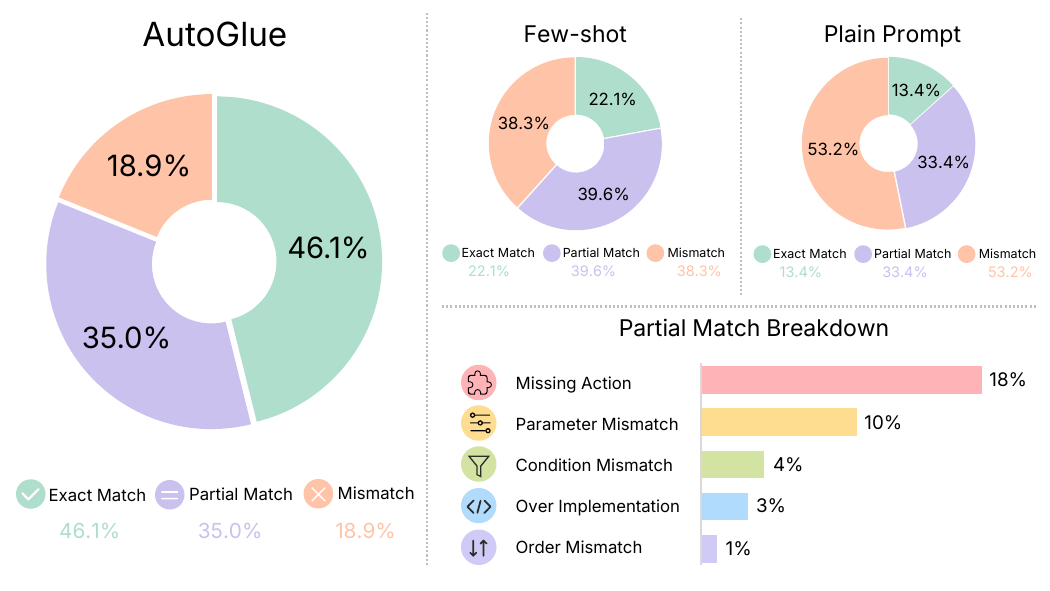}
    \captionof{figure}{Usability evaluation results and breakdown of partial match cases}
    \label{fig:llm_judge}
\end{figure*}

\phead{Results.}
\results{Nearly half of the glue code generated by \tool is directly usable, with 46.1\% of generated implementations requiring no further modification.}
Table~\ref{tab:rq2_usability} and Figure~\ref{fig:llm_judge} report the overall LLM-as-a-Judge evaluation results on 1,307 generated glue code implementations.
\tool achieves 603 Exact Match cases, corresponding to 46.1\% of all evaluated steps.
Few-shot prompting yields 289 Exact Match implementations (22.1\%), while plain prompting achieves 175 (13.4\%).
For plain prompting, the Exact Match, Partial Match, and Mismatch rates are 13.4\%, 33.4\%, and 53.2\%, respectively.
Overall, \tool achieves a 108.7\% increase over few-shot prompting and a 244.6\% increase over plain prompting in Exact Match cases.
The higher rate of direct usability suggests that \tool more reliably produces glue code that can be directly accepted and integrated into existing BDD workflows, highlighting the practical value of a purpose-built generation approach in real-world BDD projects.
In addition, \tool also lowers the proportion of unusable glue code substantially, reducing Mismatch cases to 18.9\%.


\results{Most partially correct glue code produced by \tool is close to being usable, with missing actions and parameter issues as the dominant causes.}
To better understand the remaining modification effort, we further analyze Partial Match cases by their dominant correction category.
As shown in Figure~\ref{fig:llm_judge}, the most common issue among \tool outputs is \emph{Missing Action}, accounting for 18\% of all evaluated instances.
These generated glue code typically capture the main intent of the step but omit one or more required operations.
\emph{Parameter Mismatch} accounts for an additional 10\%, where appropriate project APIs are invoked with incorrect or imprecise arguments.
In contrast, issues such as \emph{Condition Mismatch}, \emph{Over-Implementation}, and \emph{Invocation Order Mismatch} occur much less frequently.
This distribution indicates that the majority of Partial Match implementations generated by \tool would require only localized edits rather than substantial redesign, offering practical insights into how such artifacts could be efficiently corrected or refined by developers.
This breakdown also provides further structured insights for future studies on targeted correction and refinement strategies.

\results{The low frequency of over-implementation indicates that \tool adopts a conservative generation strategy that prioritizes intent alignment over speculative completeness.}
As shown in the fine-grained breakdown of Partial Match cases, \emph{Over-Implementation} occurs only rarely among the outputs of \tool, accounting for approximately 3\% of all evaluated instances.
\tool does not tend to introduce speculative logic or unnecessary project API calls in an attempt to appear plausible.
Instead, the generated glue code remains closely anchored to the semantics explicitly expressed in the step, reflecting a conservative generation strategy.
Importantly, this error profile implies that most usability issues arise from under-specification rather than misinterpretation.
The predominance of \emph{Missing Action} and \emph{Parameter Mismatch} indicates that the generated implementations are generally aligned with the correct behavioral direction, but fall short in completeness or precision.
From a practical standpoint, such errors are typically easier for developers to diagnose and correct than over-implemented logic, which may obscure intent or introduce unintended side effects.
As a result, the low frequency of over-implementation further reinforces the practical usability of \tool outputs, even in cases that are not fully correct.

\results{LLM-as-a-Judge aligns well with human evaluation at both score and usability judgment.}
We validate the LLM-as-a-Judge against human evaluation on 120 randomly sampled cases.
In the human-evaluated set, the counts for Exact, Partial, and Mismatch are 66, 26, and 28, respectively, compared with 61, 39, and 20 from the LLM judge.
The Spearman rank correlation between LLM scores and human judgments is strong ($\rho = 0.84$, $p < 10^{-30}$), indicating high correlation in relative usability ranking.
In addition, Cohen’s Kappa under a binary setting (\emph{Exact Match} vs.\ \emph{Non-Exact}) reaches $0.71$, reflecting substantial agreement in identifying directly usable glue code.
This confirms that the LLM judge provides reliable signals for both comparative usability assessment and Exact Match detection.

\smallskip
\noindent
\setlength{\fboxsep}{6pt}
\colorbox{gray!10}{%
  \parbox{\dimexpr\linewidth-3\fboxsep\relax}{%
    \textbf{RQ2 Takeaway:}
    \tool generates glue code that is largely usable in practice.
    Nearly half of the generated implementations are directly usable, and most remaining cases require only minor fixes, primarily missing actions or parameter adjustments.
    These results also show strong correlation and agreement with human evaluation.
  }
}

\subsection{RQ3: How does each component of \tool contribute to its overall effectiveness?}
\label{sec:RQ3}

\phead{Motivation.}
\tool adopts a multi-agent architecture that decomposes glue code generation into several specialized stages.
A dedicated Behavior Interpreter agent first analyzes the step to infer its intended behavior and required actions.
Supervised by the Developer agent, two retrieval sub-agents ground the generation process by retrieving relevant BDD artifacts, such as existing glue code and scenarios, as well as project code and APIs.
The retrieved contextual information is then used by the supervisor to guide the final glue code generation toward behaviorally and structurally aligned implementations.
While the full system demonstrates strong overall performance, it remains unclear how much each component contributes to glue code generation quality.
Understanding the individual impact of these design choices is essential for validating the necessity of a multi-stage, retrieval-augmented architecture and for guiding future system design and refinement.

\phead{Approach.}
We investigate how key design choices influence the effectiveness of glue code generation in \tool.
For each major component, we conduct controlled ablation experiments by removing it from the full pipeline while keeping all other components unchanged, in order to isolate its individual contribution.
This setup allows us to assess whether each component provides tangible benefits or introduces unnecessary complexity.
All variants are evaluated on the same dataset under identical experimental settings.
Generation quality is assessed using both functionality-oriented metrics (API precision, recall, and F1) and text-based similarity metrics (CodeBLEU, METEOR, and ROUGE-L).
To this end, we construct the following variants:

\begin{itemize}[leftmargin=*]
  \item \textbf{w/o Behavior Interpreter}  To evaluate the role of behavior interpretation, we remove the Behavior Interpreter agent and directly generate glue code from retrieved context. In this setting, the model no longer explicitly understands the step intent or decomposes required actions, and the input prompt is directly passed to the Developer agent.

  \item \textbf{w/o BDD Context Retriever}  To assess the importance of references from relevant BDD artifacts (e.g., other steps and glue code), we disable the BDD Context Retriever sub-agent. The Developer agent therefore lacks explicit behavioral exemplars and relies only on relevant project code to infer the intended semantics of the step.

  \item \textbf{w/o Project Context Retriever}  To examine the contribution of project code grounding, we remove the Project Context Retriever sub-agent and prevent the model from accessing relevant project code and files within the project. In this variant, glue code generation is guided by behavioral context alone, without direct exposure to project code.
\end{itemize}

\begin{table}[t]
\centering
\caption{Ablation study results on glue code generation}
\label{tab:ablation_metrics}
\begin{tabular}{lc|ccc}
\toprule
\textbf{Metric} & \textbf{\tool} & \textbf{w/o Interpreter} & \textbf{w/o BDD-Retri} & \textbf{w/o Project-Retri} \\
\midrule
API Precision   & \textbf{0.660} & 0.645 ($\downarrow$ 2.3\%)  & 0.520 ($\downarrow$ 21.2\%) & 0.552 ($\downarrow$ 16.4\%) \\
API Recall      & \textbf{0.714} & 0.654 ($\downarrow$ 8.4\%)  & 0.566 ($\downarrow$ 20.7\%) & 0.640 ($\downarrow$ 10.4\%) \\
API F1          & \textbf{0.660} & 0.622 ($\downarrow$ 5.8\%)  & 0.503 ($\downarrow$ 23.8\%) & 0.559 ($\downarrow$ 15.3\%) \\
\midrule
CodeBLEU        & \textbf{0.641} & 0.597 ($\downarrow$ 6.9\%)  & 0.519 ($\downarrow$ 19.0\%) & 0.588 ($\downarrow$ 8.3\%) \\
METEOR          & \textbf{0.655} & 0.617 ($\downarrow$ 5.8\%)  & 0.523 ($\downarrow$ 20.2\%) & 0.583 ($\downarrow$ 11.0\%) \\
ROUGE-L         & \textbf{0.614} & 0.585 ($\downarrow$ 4.7\%)  & 0.466 ($\downarrow$ 24.1\%) & 0.520 ($\downarrow$ 15.3\%) \\

\midrule
Exact Match     & \textbf{603}   & 560 ($\downarrow$ 7.1\%)    & 502 ($\downarrow$ 16.7\%)   & 474 ($\downarrow$ 21.4\%) \\

\bottomrule
\end{tabular}
\end{table}

\phead{Results.}
\results{BDD-context retrieval strongly improves overall similarity to human-written glue code.}
As shown in Table~\ref{tab:ablation_metrics}, disabling the BDD Context Retriever causes the largest degradation in similarity metrics among all components.
ROUGE-L decreases by up to 24.1\%, METEOR by 20.2\%, and CodeBLEU by 19.0\%, while API F1 drops substantially from 0.660 to 0.503.
Access to semantically related steps and existing glue code within the same project provides concrete behavioral and structural references, enabling closer alignment with established implementation conventions.
Without such references, generated glue code exhibits reduced alignment at both the token and sequence levels, leading to weaker correspondence with the intended behavior.

\results{Project-context retrieval has a significant impact on exact correctness and practical usability.}
Removing the Project Context Retriever results in the largest reduction in Exact Match cases, which drop from 603 to 474, a relative decrease of over 21\%.
At the same time, API F1 and text-based metrics decrease more moderately, by up to 15.3\%.
This pattern suggests that while BDD context supports coarse-grained behavioral realization, access to project code and APIs is essential for resolving fine-grained implementation details required for fully correct glue code.
Project-context retrieval therefore plays an important role in selecting appropriate classes, methods, and invocation patterns, influencing whether generated glue code is immediately usable without manual correction.

\results{Behavior interpretation further improves overall generation performance.}
Removing the Behavior Interpreter leads to consistent degradation across all evaluated metrics.
API Precision decreases from 0.660 to 0.645 (2.3\%), while API Recall drops from 0.714 to 0.654 (8.4\%), resulting in an overall API F1 reduction from 0.660 to 0.622.
Text-based metrics exhibit similar trends, with CodeBLEU decreasing from 0.641 to 0.597 (6.9\%) and METEOR from 0.655 to 0.617 (5.8\%).
The number of Exact Match cases is reduced from 603 to 560.
These results indicate that finer-grained inspection and interpretation of step behavior provide more precise guidance for subsequent stages of glue code generation.

\smallskip
\noindent
\setlength{\fboxsep}{6pt}
\colorbox{gray!10}{%
  \parbox{\dimexpr\linewidth-3\fboxsep\relax}{%
    \textbf{RQ3 Takeaway:}
    Each component in \tool contributes to glue code generation quality. BDD context improves similarity to human-written code, project context enables correct and usable implementations, and behavior interpretation further improves overall performance.
  }%
}


\section{Discussion}

\subsection{Efficiency of \tool}
We evaluate \tool in terms of API cost and runtime, and the results suggest it is practical.
Under a pricing model of \$0.25 per million input tokens and \$2 per million output tokens, the multi-agent pipeline costs about \$0.05 per generated step~\cite{openaiGPT5Mini}, with roughly 190k-200k input tokens and 400-1500 output tokens per step.
Cost is dominated by input tokens (context), whereas output tokens mainly come from generated glue code and tool calls.
In terms of latency, over 90\% of generations finish within 60 seconds.
Runtime is largely determined by LLM API response time (e.g., network latency and provider-side scheduling) rather than local computation.
Overall, the efficiency is sufficient for offline evaluation and developer-in-the-loop BDD workflows, where glue code generation is performed on demand rather than in performance-critical runtime paths.

\subsection{Implications}
\results{Automated glue-code generation can reduce a practical BDD bottleneck.}
Our results suggest that LLM-based agents can support one of the most labor-intensive parts of BDD: translating natural-language steps into glue code that invokes project code.
Nearly half of \tool's generated implementations are directly usable, and many remaining cases require only small fixes.
This indicates that automated generation can be useful even when it does not fully replace developers.
In practice, such tools are likely to be most valuable as project-aware assistants that draft glue code, retrieve related glue code and APIs, and help developers maintain consistency across growing BDD suites.

\results{Behavior interpretation is useful for project-aware code generation.}
The ablation results show that behavior interpretation, BDD-context retrieval, and project-context retrieval each contribute to generation quality.
This finding has implications beyond BDD glue code.
For context-dependent code generation tasks, directly prompting a model to produce code may be less reliable than first making the intended behavior explicit and then grounding implementation decisions in the local codebase.
An explicit behavior layer can reduce ambiguity in underspecified inputs, make the agent's reasoning easier to inspect, and provide a clearer interface between natural-language intent and project code.

\results{BDD suggests a broader specification-to-code paradigm.}
BDD is a useful starting point for studying how natural-language specifications can be transformed into executable code because its inputs are structured around concrete system behavior.
The success of \tool suggests that specification-to-code automation is feasible when the system can combine specification context with project code context.
This perspective is not limited to BDD.
It can inform future tools for other agile and software engineering artifacts, such as user stories, issue tickets, acceptance criteria, design notes, and lightweight task plans.
Compared with these artifacts, BDD makes behavioral intent more explicit and therefore illustrates how structure in natural-language specifications can help bridge the gap between requirements and implementation.

\subsection{Future Work}
\results{From isolated steps to scenario- and feature-level synthesis.}
\tool currently generates glue code for individual steps while using the surrounding scenario and feature as context.
An important next direction is to move from step-level generation to scenario- or feature-level synthesis.
This setting is more demanding because the generator must preserve dependencies across steps, including shared objects, fixture initialization, parameter values, background steps, and the data flow from preconditions to actions and assertions.
Generating glue code for multiple steps together could allow the system to plan a coherent execution path before emitting code, reuse variables and helper methods consistently, and avoid locally plausible implementations that conflict with other steps in the same scenario.
Future work can therefore explore structured intermediate representations, such as behavior dependency graphs or scenario-level execution plans, to guide multi-step glue code generation.

\results{Feedback-guided repair could address remaining partial and mismatch cases.}
Our usability analysis shows that many non-exact cases are close to usable, often requiring missing actions to be added or parameters to be corrected.
This error profile suggests that future systems should not treat generation as a one-shot process.
Instead, they can combine \tool's behavior-first generation with project-level feedback, including compilation errors, test failures, execution traces, exceptions, static analysis warnings, and IDE diagnostics.
Such feedback can help localize whether an error comes from step interpretation, API selection, object construction, or assertion logic, and then guide targeted repair over multiple rounds.
For harder mismatch cases, fully automatic repair may still be unreliable because the intended behavior can depend on project-specific conventions and domain knowledge.
A practical extension is therefore a human-in-the-loop workflow~\cite{gao2024taxonomy, wang2024human}, where the system presents candidate implementations, explains the main behavioral or API-level uncertainty, and lets developers select or refine the most appropriate implementation.

\results{BDD automation could support evolution and broader framework settings.}
Real-world BDD artifacts evolve together with requirements and project code.
As shown in our dataset, many scenarios are modified after their initial creation, which means glue code generation is only one part of the broader maintenance problem.
Future work can study evolution-aware BDD automation that detects when feature files, glue code, or project APIs become inconsistent and recommends corresponding updates.
This direction is especially relevant for large BDD suites, where duplicated steps, obsolete scenarios, and diverging glue implementations increase maintenance cost.
In addition, although \tool is evaluated on Java projects, its behavior-first and project-aware design is not inherently tied to Java.
Future studies can adapt and evaluate the framework across other BDD ecosystems, such as JavaScript, Ruby, Python, and PHP, to understand which parts of the approach generalize and which require framework-specific retrieval, binding, or execution support.

\section{Threats to Validity}\label{threats}

\phead{Internal threats.}
A potential internal threat is data leakage from LLM's training data.
We reduce this risk by collecting all data from open-source projects in the wild instead of curated benchmarks.
For each project, we use the latest commit available at the time of collection and treat the steps and glue code as fixed snapshots.
Another internal threat is using an LLM-as-a-Judge to assess usability.
LLM judgments can be sensitive to prompts.
We use a reference-aware setting where the judge compares generated glue code with the human-written implementation from the same project.
We test robustness by running the judge five times with the same LLM.
The Exact, Partial, and Mismatch rates vary by less than 6\% across runs.
We also compare two judge models, Gemini-2.5-Flash~\cite{google-gemini-2-5-flash} and Claude-Sonnet-4.5~\cite{anthropic-sonnet}.
The Exact Match rates differ by less than 10\%, and the Partial and Mismatch rates differ by at most 15\%.
Our human evaluation on a subset also shows high agreement (Section~\ref{sec:RQ2}).
Overall, these checks support the validity of our evaluation results.

\phead{External threats.}
The generalizability to other programming languages is a potential threat.
We focus on Java because Java glue code poses greater generation difficulty and represents one of the earliest and most widely adopted BDD standards.
Our approach does not rely on Java-specific language features. Instead, it relies on a modular separation of responsibilities that can be adapted to project-specific structures and conventions.
As a result, the conclusions drawn in this study reflect the behavior of the proposed framework under a challenging and representative BDD setting.

\section{Conclusion}
This paper presents \tool, the first automated approach for generating Java glue code directly from steps. \tool is a hierarchical multi-agent framework that separates behavior interpretation, context retrieval, and glue code generation.
We evaluated \tool on 1,307 real-world steps from eight open-source Java projects. The results show that \tool consistently outperforms prompt-based baselines in functional alignment, code similarity, and practical usability.
Our ablation results further confirm the contributions of key components of \tool. Overall, this work establishes a practical foundation for automated glue-code generation in BDD and suggests a broader behavior-to-code paradigm in which explicit intent modeling serves as a controllable layer between natural-language specifications and implementations. Future studies could build on this basis to translate other behavioral artifacts (e.g., user stories, tickets, and acceptance criteria) into executable code, extend step generation to scenario- and feature-level synthesis while maintaining consistent cross-step state and data flow, and leverage project feedback to enable more iterative, debuggable, and developer-in-the-loop automation.


\bibliographystyle{ACM-Reference-Format}
\bibliography{references}


\end{document}